\documentstyle[aps,epsbox]{revtex}
\tighten

\begin{document}

\draft


\preprint{hep-th/9911165}

\title{Brane-world solutions, standard cosmology, and dark radiation} 
\author{Shinji Mukohyama}
\address{
Department of Physics and Astronomy, 
University of Victoria\\ 
Victoria, BC, Canada V8W 3P6
}
\date{\today}

\maketitle


\begin{abstract} 

New exact solutions of brane-world cosmology are given. These
solutions include an arbitrary constant $C$, which is determined by
the geometry outside the brane and which affects the cosmological
evolution in the brane-world. If $C$ is zero, then the standard
cosmology governs the brane-world as a low-energy effective
cosmological theory. However, if $C$ is not zero, then even in
low-energy the brane-world cosmology gives predictions different from
the standard one. The difference can be understood as ``dark
radiation'', which is not real radiation but alters cosmological
evolutions. 

\end{abstract}



Superstring theories~\cite{GSW,Polchinski} have been considered as
strong candidates for the theory of everything. Among them, the
$E_8\times E_8$ heterotic superstring theory has been traditionally
considered as the most relevant one for phenomenology since $E_8$
includes the gauge group of the standard model as a subgroup. On the
other hand, recent progress on dualities between various string
theories and M-theory made it possible to investigate strong coupling
behavior of them~\cite{Witten1995}. In particular, Ho\v{r}ava and
Witten~\cite{Horava&Witten} showed that strong coupling limit of
the $E_8\times E_8$ superstring theory can be described by the
11-dimensional supergravity, which is considered as a low-energy
effective theory of M-theory, compactified on $S^1/Z_2$ of large
radius. In their scenario, the coupling constant of the heterotic
superstring theory is interpreted as radius of $S^1/Z_2$, and matter
fields are confined on two 10-dimensional timelike hypersurfaces which
correspond to fixed points of $S^1/Z_2$. Moreover, after
compactification on a Calabi-Yau manifold, the theory of matter on
each of the two fixed-point hypersurfaces, which is 4-dimensional
after compactification, is $N=1$ supersymmetric $E_8$ gauge
theory. There is evidence that over a wide range of energy scale this
5-dimensional picture
holds~\cite{Antoniadis&Quiros,Dudas&Grojean}. Thus, in this picture
our 4-dimensional universe is one of the 4-dimensional timelike
hypersurfaces, or the world volume of a 3-brane, in 5-dimensional
spacetime compactified on $S^1/Z_2$. All matter fields including those
of the standard model is confined on the 3-brane.

Although this picture seems satisfactory as a derivation of the theory
of matter fields in 4-dimension, it is expected at least for a moment
that Kaluza-Klein modes of gravitational field might spoil it. 
The large radius of $S^1/Z_2$ implies that mass gap of Kaluza-Klein
modes is small and that there appears a tower of infinite number of
light Kaluza-Klein particles, which contradicts experiments.

Recently, it was found that the above embarrassing expectation is not
correct. Randall and Sundrum~\cite{Randall&Sundrum} showed that,
for small perturbations around a flat brane-geometry, zero modes of
gravitational field, which are massless on the brane and which
correspond to gravitational field on the brane-world, are trapped 
on the brane due to the brane tension. Moreover, no Kaluza-Klein modes
which are massive in the 4-dimensional theory in the brane-world are 
strongly coupled to the gravitational zero modes and matter fields on
the brane. Therefore, Ho\v{r}ava-Witten scenario seems satisfactory as
a derivation of the theory of weak gravitational field as well as
matter fields in 4-dimension.

Now, one of the problems in string theories and M-theory is that there 
is no direct experimental or observational evidence of them. In this
respect, it seems very effective to investigate cosmological
implications of these theories, since the universe is considered to be
in very high energy at its early stage. Since the Ho\v{r}ava-Witten
scenario seems very satisfactory as explained above, many authors
investigated cosmology in the 3-brane world based on the
scenario~\cite{Brane-cosmology,SMS}. (Many complementary works have
been done on black holes~\cite{CHR,EHM}, non-linear
analysis~\cite{Chamblin&Gibbons}, gravitational force between two
bodies~\cite{Garriga&Tanaka}, and AdS/CFT
correspondence~\cite{Gubser}.) However, no global solution in
5-dimension which gives the standard cosmology in the brane world have 
been given yet. The purpose of this paper is to give it and
investigate its cosmological implications.


As a five-dimensional geometry, let us consider a metric of this form 
%
\begin{equation}
 ds_5^2 = -N(\tau,w)d\tau^2 + R(\tau,w)(dx^2+dy^2+dz^2) + dw^2. 
	\label{eqn:ansatz}
\end{equation}
We consider a negative cosmological constant, which is interpreted as
an expectation value of 3-form field in Ho\v{r}ava-Witten scenario,
and seek a general solution of the Einstein equation
$G^a_b=3k^2\delta^a_b/2$ in bulk for this ansatz. Here, $k$ is a
positive constant. It is easy to calculate the Einstein tensor for
this metric, and the result is 
%
\begin{eqnarray}
 G^{\tau}_{\tau} & = & \frac{3}{4NR^2}(2 NRR'' - \dot{R}^2),
	\label{eqn:Gtt}\\
 G^{\tau}_w & = & -\frac{G^w_{\tau}}{N} = 
	\frac{3}{4N^2R^2}(2NR\dot{R}' - N\dot{R}R' - N'R\dot{R}), 
	\label{eqn:Gtw}\\ 
 G^w_w & = & \frac{3}{4N^2R^2}
	(N^2{R'}^2 - 2 NR\ddot{R} + \dot{N}R\dot{R} + NN'RR'),
	\label{eqn:Gww}\\
 G^i_j & = & \frac{\delta^i_j}{4N^2R^2}
	(2\dot{N}R\dot{R} - 4NR\ddot{R} + 2NN'RR' - N^2{R'}^2
	+ N\dot{R}^2 + 4N^2RR'' + 2NN''R^2 - {N'}^2R^2),
	\label{eqn:Gij}
\end{eqnarray}
where dots and primes denote derivatives with respect to $\tau$ and
$w$, and $i$ and $j$ denote $x$ or $y$ or $z$. 
Since the ($\tau w$)-component of the Einstein equation, $G^{\tau}_w=0$,
becomes $(\dot{R}^2/NR)'=0$, $N(\tau,w)$ is written as 
%
\begin{equation}
 N(\tau,w) = f(\tau)\frac{\dot{R}^2(\tau,w)}{R(\tau,w)}, 
	\label{eqn:def-f}
\end{equation}
where $f(\tau)$ is a positive function of $\tau$. Substituting
Eq.~(\ref{eqn:def-f}) into Eq.~(\ref{eqn:Gtt}), the
($\tau\tau$)-component of the Einstein equation,
$G^{\tau}_{\tau}=3k^2/2$, becomes $R''-k^2R=1/2f$. Hence, $R(\tau,w)$
is written as 
%
\begin{equation}
 R(\tau,w) = \alpha(\tau)\cosh{(kw)} + \beta(\tau)\sinh{(kw)} 
	-\frac{1}{2k^2f(\tau)},
\end{equation}
where $\alpha(\tau)$ and $\beta(\tau)$ are functions of $\tau$. For
this form of $R(\tau,w)$ it is easy to confirm by using
Eq.~(\ref{eqn:Gij}) that the ($ij$)-components of the Einstein
equation, $G^i_j=3k^2\delta^i_j/2$, are satisfied. It is easy to show
that the remaining equation $G^w_w=3k^2/2$ becomes
$\partial_{\tau}(4k^4(\alpha^2-\beta^2)-1/f^2)=0$. Therefore, $f(t)$
is written as 
%
\begin{equation}
 f(\tau) =  \frac{1}{2k^2\sqrt{\alpha^2(\tau)-\beta^2(\tau)+C}},
\end{equation}
where $C$ is a constant.

Note that two functions $\alpha(\tau)$ and $\beta(\tau)$ and the
constant $C$ are arbitrary. However, these degrees of freedom should
include degrees of freedom of coordinate transformations. In fact, we
can define a new time variable $t$ and a function $a(t)$ of $t$ so
that the metric obtained above is transformed to the following form.
%
\begin{equation}
 ds_5^2 = -\frac{\psi^2(t,w)}{\varphi(t,w)}dt^2 
	+ \varphi(t,w)a^2(t)(dx^2+dy^2+dz^2) + dw^2,
	\label{eqn:sol-in-bulk}
\end{equation}
where $\psi$ and $\varphi$ are given by 
%
\begin{eqnarray}
 \psi & = & \cosh{(kw)} + 2k^{-2}(H^2+\partial_tH)(\cosh{(kw)}-1)
	\pm\frac{1+2k^{-2}(2H^2+\partial_tH)}
	{\sqrt{1+4k^{-2}H^2+Ca^{-4}}}\sinh{(kw)},
	\nonumber\\
 \varphi & = & \cosh{(kw)} + 2k^{-2}H^2(\cosh{(kw)}-1)
	\pm\sqrt{1+4k^{-2}H^2+Ca^{-4}}\sinh{(kw)},
	\label{eqn:psi-varphi}
\end{eqnarray}
(two signs of terms including $\sinh{(kw)}$ should be taken to be the
same) and $H(t)$ is determined by $a(t)$ as 
%
\begin{equation}
 H = \frac{\partial_ta}{a}.\label{eqn:Hubble}
\end{equation}
The new time variable $t$ and the function $a(t)$, which is arbitrary
for the moment, are defined by 
%
\begin{eqnarray}
 \frac{dt}{d\tau} & = & \sqrt{N(\tau,0)},\nonumber\\
 a(t) & = & \sqrt{R(\tau,0)}.
\end{eqnarray}
Therefore, the solutions of the Einstein equation
$G^a_b=3k^2\delta^a_b/2$ which have been obtained by using the ansatz
(\ref{eqn:ansatz}) are written in the form~(\ref{eqn:sol-in-bulk})
with Eqs.~(\ref{eqn:psi-varphi}-\ref{eqn:Hubble}). The function $a(t)$
is arbitrary for the moment, but the evolution equation of it will be
determined below by so called Israel's junction
condition~\cite{Israel1966} at the brane. On the other hand, the
constant $C$ will be arbitrary to the end.


Let us suppose that a $3$-brane is at $w=0$. Hence, because of the
$Z_2$-symmetry, the geometry in bulk should be invariant under parity 
transformation $w\to -w$. Thus, the 5-dimensional metric is given by
Eq.~(\ref{eqn:sol-in-bulk}) with 
%
\begin{eqnarray}
 \psi & = & \cosh{(kw)} + 2k^{-2}(H^2+\partial_tH)(\cosh{(kw)}-1)
	\pm\frac{1+2k^{-2}(2H^2+\partial_tH)}
	{\sqrt{1+4k^{-2}H^2+Ca^{-4}}}\sinh{|kw|},
	\nonumber\\
 \varphi & = & \cosh{(kw)} + 2k^{-2}H^2(\cosh{(kw)}-1)
	\pm\sqrt{1+4k^{-2}H^2+Ca^{-4}}\sinh{|kw|}.
\end{eqnarray}
(As in Eq.~(\ref{eqn:psi-varphi}), two signs of terms including
$\sinh{|kw|}$ should be taken to be the same.) 
Note that the induced metric on the brane at $w=0$ is 
%
\begin{equation}
 q_{\mu\nu}dx^{\mu}dx^{\nu} = - dt^2 + a^2(t)(dx^2+dy^2+dz^2), 
\end{equation}
and, thus, is the flat FRW metric with the scale factor $a(t)$.

Now, in order to give evolution equations of $a(t)$, we use so
called Israel's junction condition~\cite{Israel1966}:
%
\begin{eqnarray}
 \left[q_{\mu\nu}\right] & = & 0,\label{eqn:junction1}\\
 \left[K_{\mu\nu}-Kq_{\mu\nu}\right] & = & -\kappa_5^2 S_{\mu\nu},
	\label{eqn:junction2}
\end{eqnarray}
where $[X]$ denotes $\lim_{w\to +0}X-\lim_{w\to -0}X$, $S_{\mu\nu}$ is 
the surface energy-momentum tensor on the brane, and the extrinsic 
curvature $K_{\mu\nu}$ of the constant-$w$ hypersurface and its trace
$K$ are defined by 
$K_{\mu\nu}\equiv
q_{\mu}^{\alpha}q_{\nu}^{\beta}\nabla^{(5)}_{\alpha}n_{\beta}$ and 
$K\equiv K^{\mu}_{\mu}$. Here, $\nabla^{(5)}_{\alpha}$ is the
five-dimensional covariant derivative with respect to the metric we
have obtained and $n^{\mu}$ is the unit normal to the constant-$w$
hypersurface given by $n^{\mu}\partial_{\mu}=\partial_w$. Since
Eq.~(\ref{eqn:junction1}), which represents a gauge condition, is
satisfied automatically for our metric, we shall concentrate on
Eq.~(\ref{eqn:junction2}).

We assume that the surface energy-momentum tensor is a sum of a
surface tension term and a perfect fluid term: 
%
\begin{equation}
 S_{\mu\nu} = -\lambda q_{\mu\nu} + T_{\mu\nu},\nonumber\\
\end{equation}
where 
%
\begin{equation}
 T^{\mu}_{\nu} = \left(\begin{array}{cccc}
	-\rho(t) & 0 & 0 & 0 \\
	0 & p(t) & 0 & 0 \\
	0 & 0 & p(t) & 0 \\
	0 & 0 & 0 & p(t) 
	\end{array}\right).
\end{equation}
As in Refs.~\cite{Randall&Sundrum}, 
the surface tension is assumed to be related to the bulk cosmological
term as 
%
\begin{equation}
 \lambda = \mp 3\kappa_5^{-2}k. \label{eqn:lambda}
\end{equation}
This relation arise in the five-dimensional effective theory of the
Ho\v{r}ava-Witten scenario~\cite{Horava&Witten,LOSW}. It would be
worth while mentioning that deviation from Eq.~(\ref{eqn:lambda}) will 
produce non-zero cosmological constant in the brane-world.

Since $K^{\mu}_{\nu}-Kq^{\mu}_{\nu}$ is given by
%
\begin{eqnarray}
 K^{\tau}_{\tau} - K & = & -\frac{3}{2}\partial_w\ln\varphi,
	\nonumber\\
 K^i_j - K\delta^i_j & = & -\frac{1}{2}(
	2\partial_w\ln\psi+\partial_w\ln\varphi )\delta^i_j,
\end{eqnarray}
the left hand side of Eq.~(\ref{eqn:junction2}) is calculated as
%
\begin{eqnarray}
 \left[K^{\tau}_{\tau} - K\right] & = & 
	\mp 3k \sqrt{1+4k^{-2}H^2+Ca^{-4}},\nonumber\\
 \left[K^i_j - K\delta^i_j\right] & = & 
	\mp k \frac{3+4k^{-2}(3H^2+\partial_tH)+Ca^{-4}}
	{\sqrt{1+4k^{-2}H^2+Ca^{-4}}}.
\end{eqnarray}
Thus, Eq.~(\ref{eqn:junction2}) becomes 
%
\begin{eqnarray}
 3k\left(\sqrt{1+4k^{-2}H^2+Ca^{-4}}-1\right) & = &
	\mp\kappa_5^2\rho,\nonumber\\
 3k\left(\frac{3+4k^{-2}(3H^2+\partial_tH)+Ca^{-4}}
	{3\sqrt{1+4k^{-2}H^2+Ca^{-4}}} - 1\right) & = & 
	\pm\kappa_5^2p.		\label{eqn:cosmological-eq}
\end{eqnarray}
These are the cosmological equations on the brane-world. These
equations combined with an equation of state determines $a(t)$,
$\rho(t)$ and $p(t)$ uniquely as functions of time, provided that
suitable initial conditions are given. The constant $C$ is still
arbitrary and its physical meaning as ``dark radiation'' will be
discussed below.

The undetermined signs in the above equations will be determined by 
considering a low-energy limit as follows.
In order to seek low-energy effective cosmological equations, we
should consider the case in which $k^{-2}H^2\ll 1$ and
$Ca^{-4}\ll 1$. In this case, Eqs.~(\ref{eqn:cosmological-eq}) become 
%
\begin{eqnarray}
 H^2 & = & \mp \frac{8\pi G_N}{3}\rho -\frac{Ck^2}{4a^4},
	\nonumber\\
 \frac{\partial_t^2a}{a} & = & \pm \frac{4\pi G_N}{3}(\rho+3p)
	+\frac{Ck^2}{4a^4}, \label{eqn:Friedman-eq}
\end{eqnarray}
where the Newton's constant $G_N$ is given by
%
\begin{equation}
 G_N = \frac{k\kappa_5^2}{16\pi}.
\end{equation}
Eqs.~(\ref{eqn:Friedman-eq}) are the same as equations in the standard
cosmology, provided that the lower signs are taken and the terms
linear in $a^{-4}$ are absorbed in $\rho$ and $p$. Therefore, the
lower signs should be taken in all equations.

If the arbitrary constant $C$ is zero, then the low-energy evolution
equations of $a(t)$ completely agree with equations in the standard 
cosmology. However, if $C$ is not zero, then new terms appear in the
evolution equations. We shall now investigate how these terms can be
interpreted. Eqs.~(\ref{eqn:Friedman-eq}) can be written as 
%
\begin{eqnarray}
 H^2 & = &  \frac{8\pi G_N}{3}\rho_{eff},\nonumber\\
 \frac{\partial_t^2a}{a} & = & - \frac{4\pi G_N}{3}
	(\rho_{eff}+3p_{eff}),
\end{eqnarray}
where 
%
\begin{eqnarray}
 \rho_{eff} & = & \rho - \frac{3Ck}{2\kappa_5^2}\frac{1}{a^4},
	\nonumber\\
 p_{eff} & = & p - \frac{1}{3}\frac{3Ck}{2\kappa_5^2}\frac{1}{a^4}.
\end{eqnarray}
Thus, the terms linear in $a^{-4}$ can be considered as a shift of
radiation density. Since non-zero value of $C$ does not
implies real existence of radiation but changes cosmological evolution
equations, it can be understood as ``dark radiation''. Here, note that
the constant $C$ can be both positive and negative, depending on the
geometry in bulk. To be precise, the term proportional to $C$ can be
understood as the 'electric' part of the $5$-dimensional Weyl tensor
in the general frame work developed in Ref.~\cite{SMS}. Moreover, it
can be shown that if $C$ is positive then there appears a naked
singularity in $5$-dimension~\cite{MSMS}. Thus, the constant $C$
should be zero or negative, which implies that energy density of the
dark radiation should be zero or positive.


In summary, we have obtained new exact solutions of brane-world
cosmology. These solutions include an arbitrary constant $C$. The
constant $C$ is determined by the geometry outside the brane and
affects the cosmological evolution in the brane-world. If $C$ is zero,
then the standard cosmology governs the brane-world as a low-energy
effective cosmological theory. However, if $C$ is not zero, then even
in low-energy the brane-world cosmology gives predictions different
from the standard one. The difference can be understood as ``dark
radiation'', which is not real radiation but alters cosmological
evolutions. Density of the dark radiation should be zero or positive.

{\bf Note added in proof}: After this paper had been submitted, the
author became aware of the works~\cite{Kraus,BDEL,Vollic,Ida}, in
which similar solutions were obtained.

\vspace{1cm}

The author would like to thank Professor W. Israel for helpful
discussions and continuing encouragement. The author is supported by
the CITA National Fellowship and the NSERC operating research grant.


\end{document}